# Observation of neutrals carrying ion-acoustic wave momentum in partially ionized plasma

Meenakshee Sharma[1,2], A.D. Patel[1], Zubin Shaikh[1,3], N. Ramasubramanian[1,2], R. Ganesh[1,2], P. K. Chattopadhayay[1,2], and Y.C. Saxena[1]

[1]Institute for Plasma Research, Gandhinagar, India
[2] Homi Bhabha National Institute, Anushaktinagar, Mumbai, India
[3]Department of Physics, Saurashtra University, Rajkot, India

*Email: meenakshee.26@gmail.com*

## Abstract

An experimental study of Ion Acoustic (IA) wave propagation is performed to investigate the effect of neutral density for argon plasma in an unmagnetized linear plasma device. The neutral density is varied by changing the neutral pressure, which in turn allows the change in ion-neutral, and electron-neutral collision mean free path. The collisions of plasma species with neutrals are found to modify the IA wave characteristics such as the wave amplitude, velocity, and propagation length. Unlike the earlier reported work where neutrals tend to heavily damp IA wave in the frequency regime $\omega < \nu_{in}$ (where $\omega$ is ion-acoustic mode frequency and $\nu_{in}$ is ion-neutral collision frequency), the experimental study of IA wave presented in this paper suggests that the collisions support the wave to propagate for longer distances as the neutral pressure increases. A simple analytical model is shown to qualitatively support the experimental findings.

# I. Introduction

Ion Acoustic (IA) waves are low-frequency electrostatic, longitudinal waves in plasma. In 1929, Lewi Tonks and Irving Langmuir provided details of their experiments on oscillations in plasmas as well as a derivation of the phase velocity of IA waves associated with the oscillations [1]. In 1933, J. J. Thomson, provided a derivation of IA waves from ion fluid equations and the Boltzmann relation for electrons, and it has been the basis of several fluid model derivations of IA waves ever since. The first experimental measurements of standing IA waves were reported by Revans [2] in 1933, whereas the observation of propagating IA waves took another 33 years, and it was reported by Wong et al. [3] in the Q machine in 1962. The IA wave has also been observed in astrophysical plasmas, such as solar wind [4], Saturn ring [5], Comets, Neutrinos [6], Magnetosphere as well as in fusion plasmas.

IA wave frequency, wavelength, and phase velocity data have been the basis of valuable plasma diagnostics [7,8]. In electron-ion plasmas, IA wave phase velocity provides the electron temperature, $T_e$. The phase velocity of IA waves has also been used to measure the relative ion concentrations of two ion species in plasma [9]. Negative and positive ion plasmas are other example of two ion species plasmas for which IA waves provides ion number density ratios [10–13]. In the part, IA phase velocity measurements were shown to provide a technique to determine the drift velocity in sheath-presheath plasma [9,11,14,15].

Many decades after Langmuir and Tonk [1] first identified this IA wave, it is still a rich and exciting subject. Excitation of IA wave and their propagation in plasma has been studied by many researchers in past decades [8,16–24]. Several phenomena of IA wave, such as diffraction [25–27], reflection [28], and wave damping [3,29–33], have been topics of attraction for many researchers over the years. However, very few experimental works have been reported on the interactions of neutral particles with plasma species in IA wave propagation [34,35].

Neutral particles are important and integral component of laboratory plasma and play a vital role in the characteristics of the plasma. In partially ionized plasmas (where, ionization < 1%) neutrals are known to modify the plasma characteristics. Even in high-temperature plasmas, such as Tokamak, very small amount of neutrals in the edge is known to affect the overall equilibrium configuration and dynamical behavior of plasma, leading to L to H transition triggered by radial electric field generated via neutral particle ionization [36,37]. Partially ionized plasma is not uncommon in space, planetary ring system, asteroid zone, comets, magnetosphere, and the low-altitude ionosphere are few examples of partially ionized plasma in the space environment.

The neutral particles move undisturbed until they make a collision with another plasma specie, and these collisions affect the particle's motion. The influence of the neutral collision on IA wave is very intervening, and the details are not fully explored yet in experiments.

In earlier work, the studies on the effect of neutral collision on the IA wave propagation support the conventional idea that neutrals take away the momentum of wave and render heavy wave damping in collisional regime of plasma [34–36]. In our experiment, the IA wave is excited in the quiescent plasma regime of the linear cylindrical plasma device [42]. The basic characteristics of IA wave are studied in detail. To further explore the effect of neutral collisions on IA wave, experiments are performed for varying neutral pressure. This experimental observation brings rather surprising and counterintuitive results in contrast to earlier reported work [34–36]. For example, in our experiments we observe that for certain collisional regimes, the neutrals impart momentum to ions and as a result enhance the propagation length scale of IA wave. This phenomenon is not often observed in IA wave experiments. In this article a detail study of effect of neutral collision on IA wave to support the enhancement in wave propagation length scale is reported.

The paper is organized as follows; the experimental setup is described in Section 2, in Section 3 the basic plasma parameter is discussed, the typical characteristic of IA wave is discussed in Section, and in Section 5 the effect of neutral on IA wave is discussed followed by summary and conclusion.

## II. Experimental Setup

The ion-acoustic wave propagation study is performed in a cylindrical **M**ulti-pole line cusp **P**lasma **D**evice (MPD) [38]. This device has facility to operate with or without magnetic field and for present experiments the MPD is used as an unmagnetized linear plasma device. The schematic of the experimental setup shown in Figure 1. The device consists of main vacuum vessel of 150 cm long stainless-steel cylinder with 6 mm wall thickness and 40 cm diameter. The base pressure of $2 \times 10^{-6}$ mbar is achieved through a pumping system with a combination of Rotary and Turbo-molecular pump. The experiments are performed for Argon gas at different fill pressure varying from $20 \times 10^{-4}$ mbar to $2 \times 10^{-4}$ mbar.

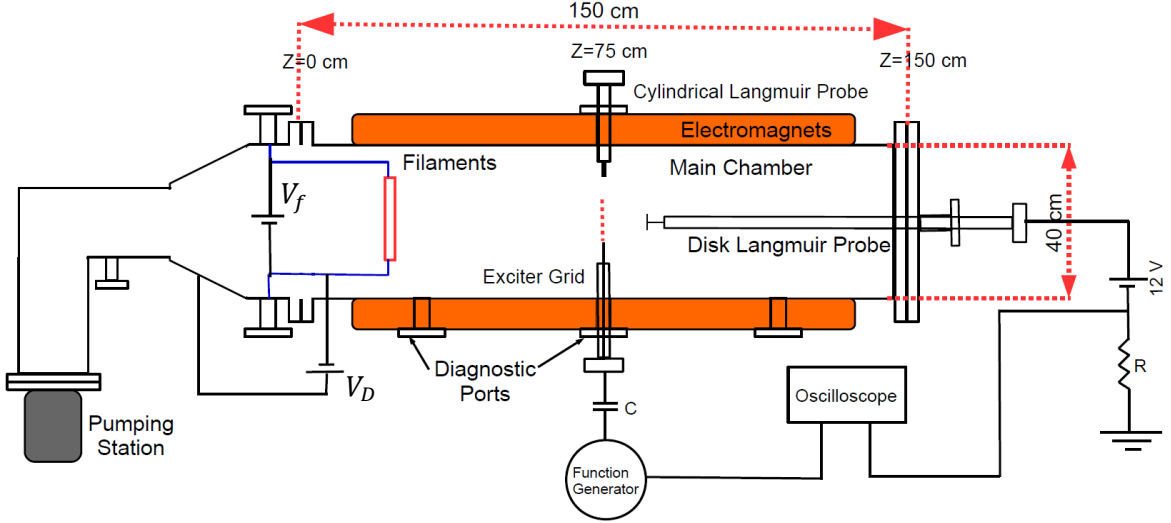

*Figure 1: Schematic of the Experimental Device and exciter circuit*

The Argon plasma is produced by an array of 5 tungsten filaments each having 80 mm length and 0.5 mm diameter. These filaments are connected in parallel for the electron emission with equal energy distribution. A floating power supply of 15V/500A is used to heat the filaments with a typical filament current of 80 -100 A, such that 16-20 A current passes through per filament. The electrons emitted from filaments get accelerated through applied discharge voltage of -50 V with respect to grounded chamber wall. With above configuration discharge current of 2 to 6 A is achieved. The accelerated electrons produce plasma of density $\sim 10^{15}$-$10^{16}$ m$^{-3}$ by impact ionization and electron temperature, $T_e$ =3-5 eV. The plasma parameters such as plasma density ($n_e$), electron temperature ($T_e$), plasma potential ($\phi_p$), and floating potential ($\phi_f$) are measured by using Langmuir probe (LP) [39].

Excitation of IA wave through a grid placed inside the plasma column is the most extensively used technique, we follow the same experimental method to study IA wave in the present work. A circular grid is installed at $R = 0, Z = 75$ cm, at centre of the device, shown in Figure 1. This exciter is stainless steel (SS304) mesh and has 50 mm diameter made of a wire of diameter 0.08 mm and 0.3 mm aperture, with 85% transparency. A time varying sinusoidal potential of 2 $V_{p-p}$ (peak to peak voltage) is applied to the exciter grid to perturb the plasma density using a function generator. The plasma response to applied perturbation is detected by an axially movable disk Langmuir probe (receiver probe) of 8 mm diameter at R=0 cm. The circuit used for excitation and detection of IA wave is shown in Fig 1. The receiver probe is biased near the local plasma potential ($\sim 12\, V$) to measure electron saturation current. The measurement of electron saturation current (IA wave signal) is performed using a 12 V battery based circuit. The data is acquired using an 8-bit digital oscilloscope at a sampling rate of 250 MSa/sec, and stored for further analysis.

The IA wave is launched below ion plasma frequency ($\omega_{pi}$) to avoid the damping due to resonance with ion plasma frquency. Identification of the frequency regime for excitation of such wave requires information of basic plasma parameters. The detail of experimental measurement of plasma parameters is discussed in the following section.

### III. Basic Plasma Parameters

The mean plasma parameters are measured using I-V characteristic of single Langmuir probe (LP). A sweep voltage ramp of -80 V to 25 V is applied on the Langmuir probe, to obtain I-V characteristics using I to V converter circuit. The regime below -40 V of this I-V characteristic is used to measure ion saturation current ($I_{isat}$). The electron current is obtained by subtracting $I_{isat}$ from total probe current. The logarithmic slope of electron current provide electron temperature. The plasma density is calculated by using ion-saturation current and electron temperature. The measured electron temperature ($T_e$) and plasma density ($n$) with different Argon gas pressure is shown in figure 2.

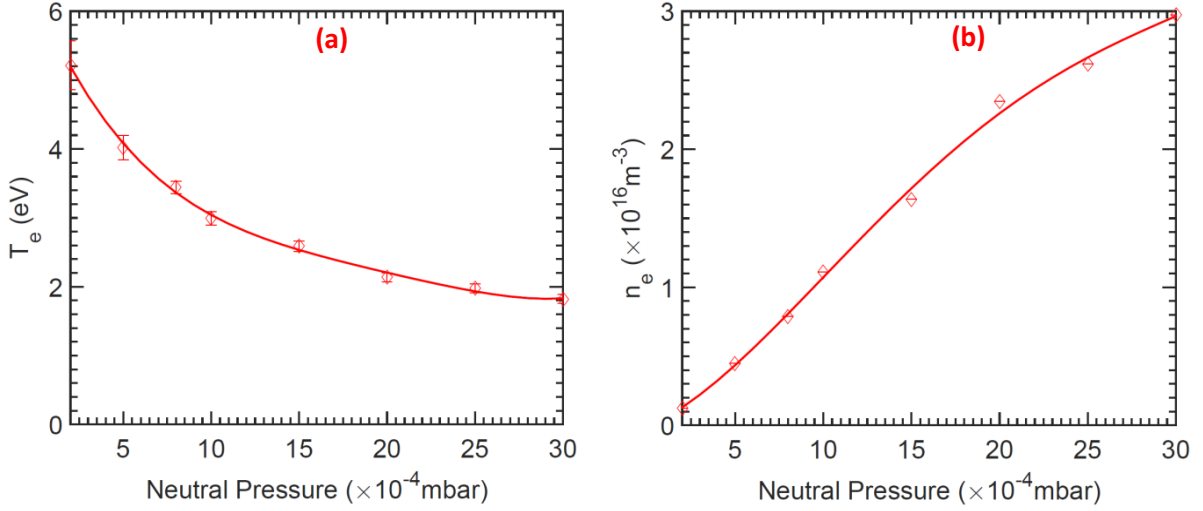

*Figure 2: Variation of (a) electron temperature, $T_e$, and (b) plasma density, $n$ at $R = 0$ cm, as a function of neutral pressure for -50 V discharge voltage.*

To identify the effect of neutral density on the evolution of plasma parameters Argon gas pressure is varied from $2 \times 10^{-4} - 30 \times 10^{-4}$ mbar. The plasma parameters are measured at R=0 cm, and Z=75 cm centre of the cylindrical device, with -50 V discharge voltage. As the neutral pressure (density) is increased from $2 \times 10^{-4}$ mbar to $30 \times 10^{-4}$ mbar the plasma density is found to increase from $1.5 \times 10^{15} - 3 \times 10^{16}$ m$^{-3}$ and electron temperature is found to decrease from ~6 eV to ~2 eV for -50 V discharge voltage. The electron-neutral collision mean free path, $\lambda = 1/n_n \sigma_{en}$, decreases with increase in neutral density, electrons suffer more collisions before they hit the wall of the grounded chamber, and ionization increases as a result. Hence the plasma density enhances with increase in neutral density, and it leads to a decrease in electron temperature. For above background plasma condition, the ion plasma frequency $\omega_{pi}$ varies from $10^6 - 10^8$ Hz. Therefore, in this work the perturbation frequency to study the IA wave propagation is chosen to be below $10^6$ Hz.

## IV. Typical features of ion-acoustic wave

A potential perturbation of $2V_{p-p}$ is launched in the plasma for $30 \times 10^{-4}$ mbar neutral pressure. The nature of perturbation is sinusoidal with a frequency of 100 kHz which is well below the ion plasma frequency ($\omega_{pi}/2\pi \sim 10^6 \, Hz$). The propagation of this potential perturbation is recorded at different axial locations using a movable receiver probe away from the exciter grid. The separation between exciter grid and receiver probe is denoted as *d*. The typical experimental observation of wave propagation is shown in figure 3(a). It shows the temporal evolution of normalized plasma density fluctuation ($\delta n/n$) measured at different axial separation, *d* from exciter grid. Experimental observation shows that received wave amplitude ($\delta n/n$) decreases as the wave propagates spatially in the plasma column. The received wave signal has three peaks and the perturbation signal has only two peaks, so the phenomena can be described as plasma response to the perturbation as follows. During the rise phase of perturbation signal (from A to B) a change in local potential is excited in plasma and as plasma response peak 1 is observed, similarly during the long fall phase of perturbation signal (from B to C), plasma response to the change in local potential excites the second peak, and on the same way the third peak is also observed. The probe signal shows the local measurement of plasma response to the given perturbation, hence 3 peaks are observed in IA wave signal. When the IA wave is propagating in space, it demonstrates time delay at every 2 cm separation. The velocity of IA wave can be calculated using the information of time delay of wave and distance it has traveled during this time.

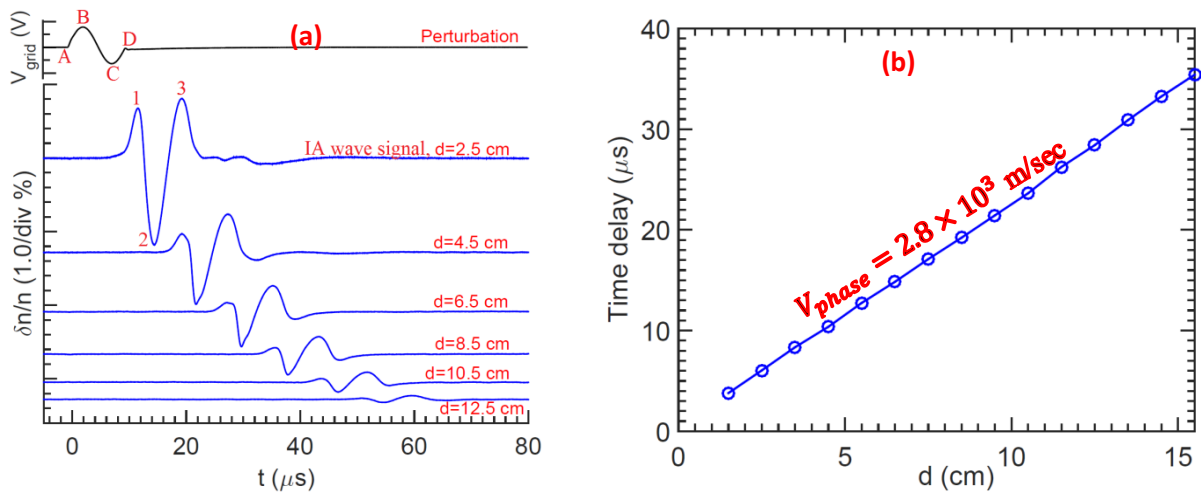

*Figure 3: (a) Shows a temporal evolution of density perturbation ($\delta n/n$) at different axial locations, and (b) time delay of the first peak of wave as the receiver probe moves away from the exciter grid along the z-axis for -50 V discharge voltage, $30 \times 10^{-4}$ mbar neutral pressure, and 100 kHz perturbation frequency, here d is the distance between exciter grid and receiver probe.*

Figure 3(b) shows the time delay of the first peak of wave versus the position of receiver probe which shows a linear variation. The phase velocity of IA wave propagation is $2.8 \times 10^3$ m/sec measured by the time of flight method of first peak as shown in figure 3(b).

It is compared with theoretically calculated ion-acoustic velocity for Argon plasma, $C_s = 2 \pm 0.3 \times 10^3$ m/sec and matches well ($C_s = \sqrt{K_B T_e / m_i}$ ).

Further experiments are performed to obtain the wave features such as $\omega - k$ relation, plasma density and potential phase relation, and power spectrum. Figure 4(a) shows the dispersion relation for the mentioned experimental observations. In this figure, the propagation frequency of wave is same as the perturbation frequency. The wavenumber, $k$ is governed by plasma conditions. In this experiment, the wave frequency is varied externally. The wavenumber $k$ is calculated using $\omega$ and measured phase velocity, $v_{phase}$, $\omega/k = v_{phase}$. The straight line shows a constant velocity nature of wave with change in perturbation frequency, and it is one of the characteristics of the IA wave [35,40].

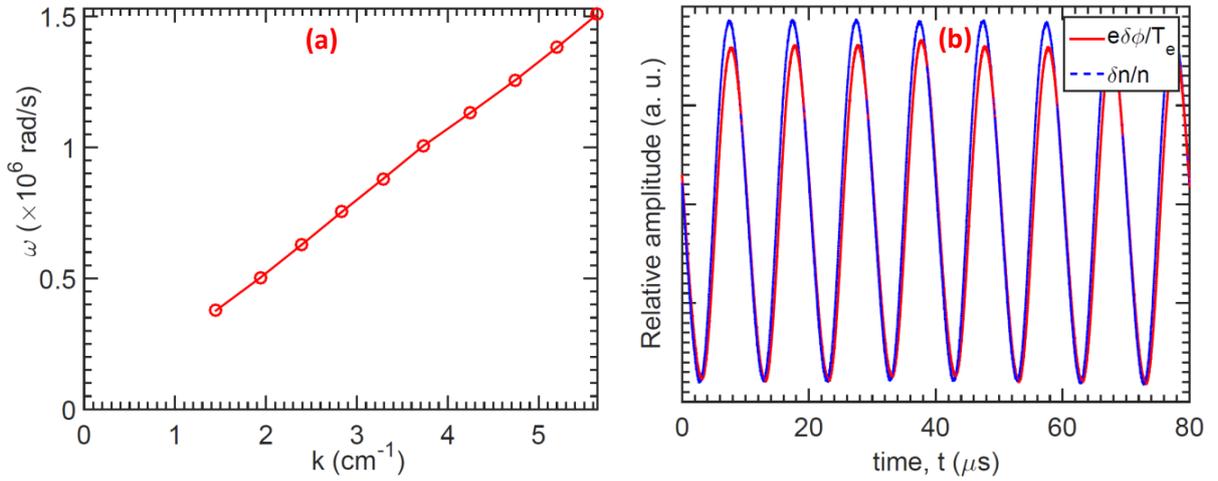

*Figure 4: (a) Experimentally obtained dispersion curve ($\omega$ vs $k$) for $\omega_{pi} = 2 \times 10^6$ Hz and $\lambda_{De} = 0.08$ mm, and (b) Temporal evolution of normalized plasma density, n and potential, $\phi$ fluctuations.*

For IA wave the relation between plasma density and potential fluctuations, for linear regime of wave, is given by

$$\frac{\delta n}{n} \sim \frac{e \delta \phi}{T_e}$$

The density fluctuation ($\delta n$) is normalized by mean plasma density ($n$), and potential fluctuation ($\delta \phi$) is normalized by mean electron temperature ($T_e$). Figure 4(b) shows the temporal characteristic of $e\phi/T_e$, and $\delta n/n$. As noted, the temporal evolution of both the quantities is in the same phase and has same amplitude.

The power spectral analysis is performed to observe the effect of the external perturbation on background plasma fluctuations. The temporal evolution of fluctuation in plasma density for two plasma conditions, without and with the external perturbation, is shown in figure 5(a). The level of fluctuation is low for unperturbed plasma, which is ≤ 0.1% can be considered as quiescent plasma. The normalized plasma density fluctuations are observed to enhance in the presence of external potential perturbation as shown in figure 5(a). The power spectrum of both signals is shown in figure 5(b). For quiescent (unperturbed) plasma, absence of any recognizable peak demonstrates the non-existence of power coupling in significant mode. As the perturbation of 100 kHz is excited in the plasma the frequency spectra show a sharp peak at100 kHz along with its harmonics.

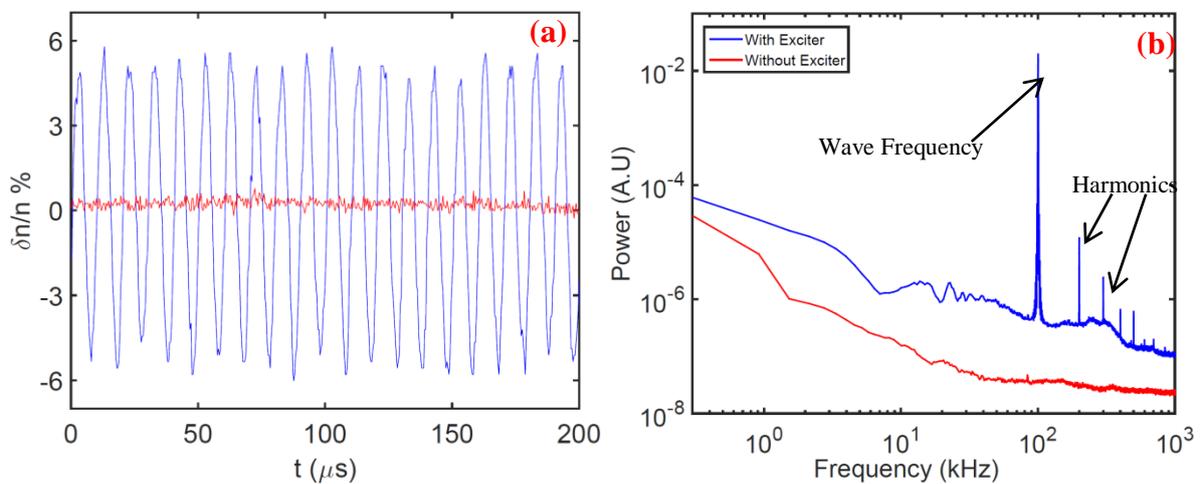

*Figure 5: (a) Time evolution of normalized density fluctuation, $\boldsymbol{\delta n/n}$, the red trace shows the density fluctuation of quiescent plasma, and blue trace shows the density fluctuation when perturbation of $2\boldsymbol{V_{p-p}}$ and 100 kHz is excited in plasma. (b) shows the power spectrum of both the signals the lower trace (red) is for quiescent plasma and the upper trace is when the perturbation of $2\boldsymbol{V_{p-p}}$ and 100 kHz is excited in plasma. The signal is recorded at d=4.5 cm.*

From figure 5(b) few more observations can be made such as (i) the propagation frequency of the wave is same as the perturbation frequency; hence, we are exciting the known frequency in plasma, (ii) in presence of perturbation, the maximum power is coupled to the excited IA wave. Hence only the IA wave is excited in plasma with the launched sinusoidal signal, and it is not exciting any other wave in the plasma.

As observed in figure 3, the wave amplitude ($\delta n/n$) decreases spatially when wave propagates in plasma. The spatial variation of observable propagation length is shown in figure 6. The maximum distance away from the exciter, up-to which the wave amplitude can be measured within the resolution of the oscilloscope, is observable propagation length of IA wave for this experimental condition. The signal amplitude measured at all locations is normalized with the amplitude of the signal at 2.5 cm, along the wave propagation axis (Z-axis of cylindrical device).

Several possible mechanisms can contribute to wave damping phenomena in plasma, such as Landau damping [41,42], plasma density gradient [43] driven damping along the wave propagation axis, and collisions of plasma species (ions, electrons, and neutrals) [32–34].

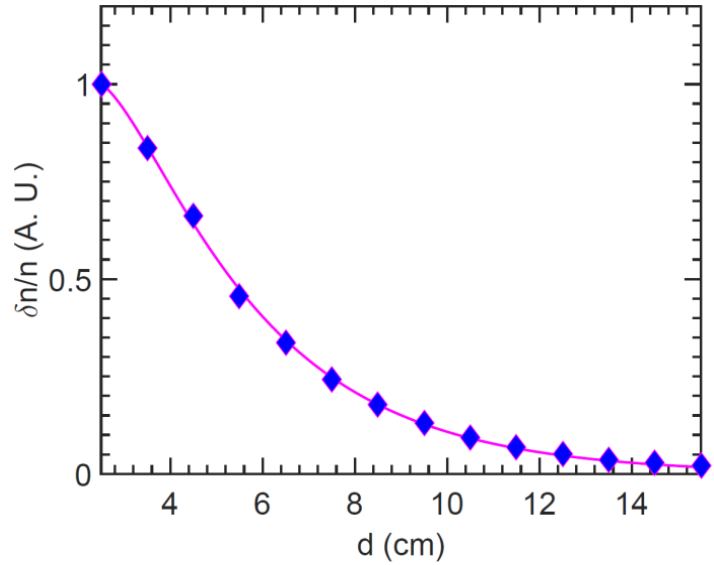

*Figure 6: The spatial variation of $\delta n/n$ or wave amplitude for $30 \times 10^{-4}$ mbar neutral pressure. The diamond-shaped markers are experimental data, and the solid lines are an exponential fit to the experimental data.*

Landau damping is one of the most important and widely studied phenomena predicted by collisionless plasma theory [30,41,42,44,45] and first observed by Wong in 1962 [3] for IA wave. This dissipationless damping takes place when the energy of the plasma wave transfers to ions moving at nearly equal to the phase velocity of the wave. In our experimental conditions, $T_e > T_i$, for this the ion thermal speed, $V_{thi} = 700$ m/sec much less than the wave phase velocity, $\omega/k = 2.8 \times 10^3$ m/sec. Hence, it may be interpreted that Landau damping, may not be the reason for observed damping.

The gradient in plasma density along the wave propagation axis can be a source to make the amplitude decrease spatially as discussed by Doucet et al. [43]. To obtain more insight into the spatial decrease in the wave amplitude, we look for the axial (along the propagation axis of the wave) variation of plasma parameters. The axial profile of plasma parameters is measured at centre of the device, R=0 cm, for $5 \times 10^{-4}$ mbar neutral pressure. Electron temperature $T_e$ is nearly constant over the axial length and plasma density $n$ has a gradient as shown in figure 7(a). The scale length of density is (~30 cm) constant upto Z=95 cm as shown in figure 7(b). The gradient scale length of density is higher than the wavelength of IA wave ($L_n > \lambda_{wave} \sim 6 - 3$ cm). Hence the gradient in plasma density cannot be the reason for the spatial wave damping.

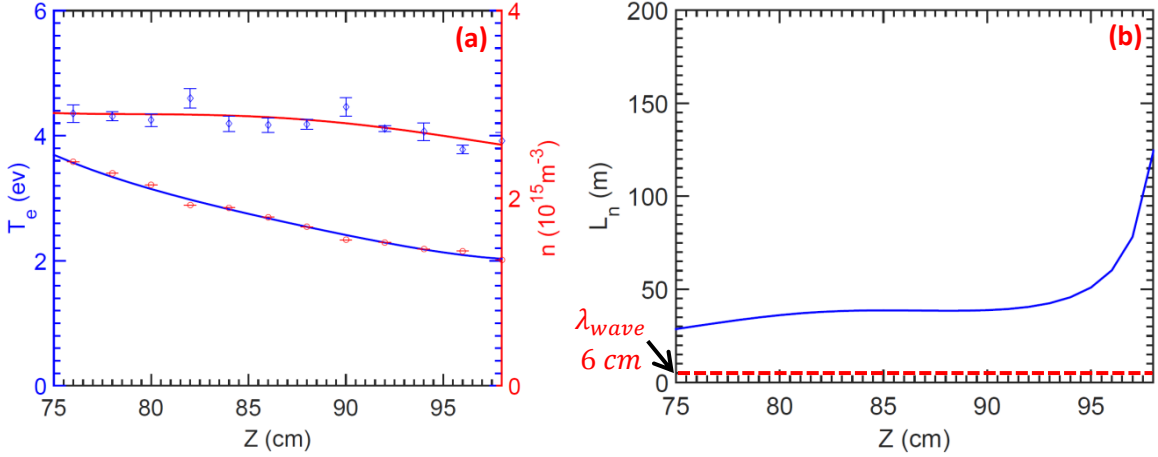

*Figure 7: Axial variation of (a) electron temperature, $T_e$, and plasma density, n, and (b) plasma density scale length, $L_n$ the dotted line shows value of IA wavelength ($\lambda_{wave}$) for -50 V discharge voltage and neutral pressure $5 \times 10^{-4}$ mbar. The exciter grid is placed at Z=75 cm.*

Collision between the plasma species can take the momentum away from the wave. In partially ionized plasma, neutrals represent a barrier for electron and ion motion in a wave field.

The effect of neutral collisions on the IA wave is explored as described in the following section. The experiments on IA wave presented in the next section are performed at 100 kHz perturbation frequency and $2V_{p-p}$ perturbation voltage unless specifically stated.

## V.     Effect of neutral density on IA wave

In partially ionized plasma, the presence of neutral is found to modify the IA wave propagation. A detailed experiment is performed for various neutral densities to quantify the effect of neutral density on the IA wave in LPD. The normalized wave amplitude ($\delta n/n$) is shown in figure 8, for several values of Argon gas pressure. The pressure is a representation of neutral density in experimental studies. The wave amplitude shown in the figure is measured at d=2.5 cm, with various pressure values. The horizontal position of each set, obtained for different neutral density, is displaced to avoid overlapping in the figure. The wave amplitude increases as the neutral pressure increases from $2 \times 10^{-4}$ mbar to $30 \times 10^{-4}$ mbar.

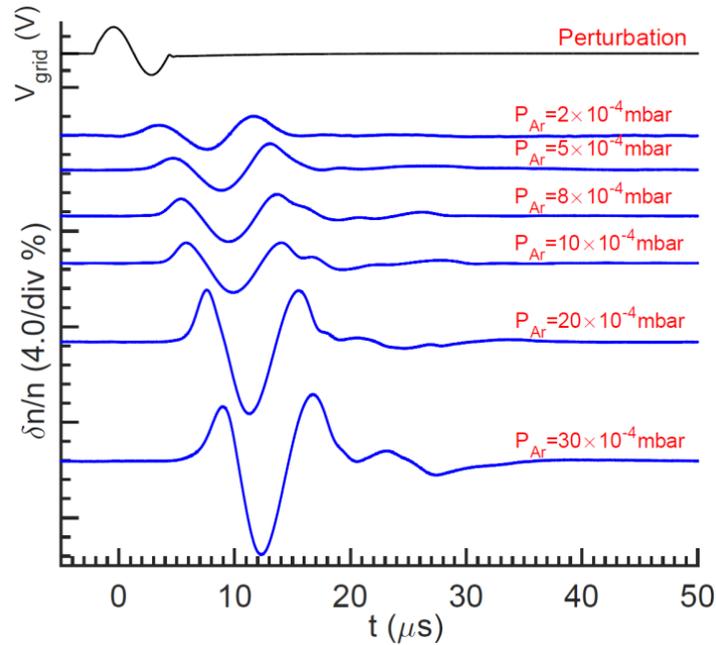

*Figure 8: Propagation of IA wave for various neutral pressures, the signal of electron saturation current is measured at d=2.5 cm for -50 V discharge voltage, here d is the distance between the exciter grid and receiver probe.*

The spatial variation of wave amplitude measured at d=2.5 cm and the effect of change in neutral density on this spatial variation is discussed further.

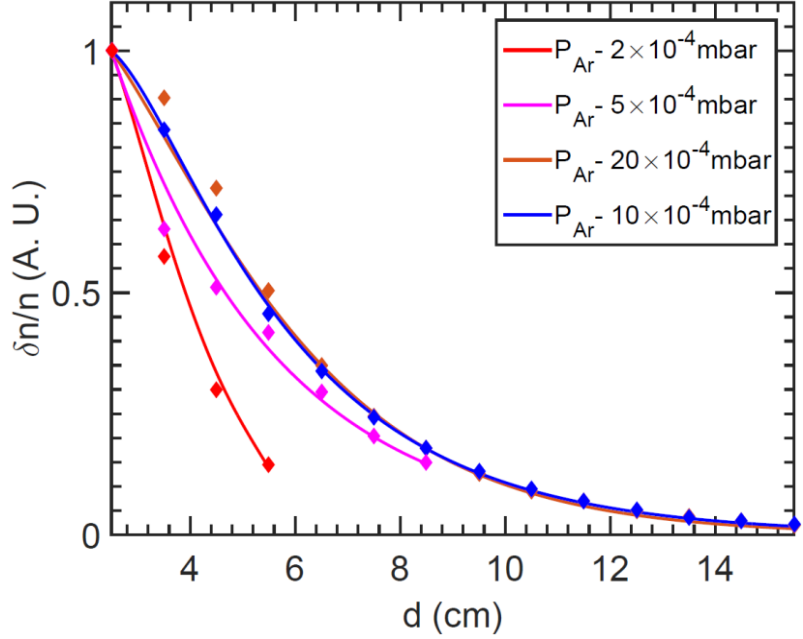

*Figure 9: The spatial variation of $\delta n/n$ or wave amplitude for various neutral pressures. The diamond-shaped markers are experimental data, and the solid lines are an exponential fit to the experimental data.*

The observable propagation length by varying the neutral pressure within the experimental limit, from $2 \times 10^{-4}$ mbar to $20 \times 10^{-4}$ mbar is presented to show the effect of neutral density on wave propagation [Figure 9]. The signal amplitude measured at all locations is normalized with the amplitude of the signal measured at d = 2.5 cm. It is observed from figure 9 that as the neutral pressure increases, the wave propagation length increases from 5 cm to 15 cm. Also one can infer from this figure that wave amplitude decreases sharply for low neutral pressure. The observations suggest that wave propagates for longer distances for high neutral pressure.

The increase in observed propagation length with an increase in neutral density is in contrast to the earlier reported work [34,35]. As the neutral density increases, the collision of neutrals with plasma species also increases, as shown in table 1. As discussed in the literature, the propagation length should decrease with increase in neutral density. In a theoretical work, Vranjes et al. [46] found a region where neutrals do support wave propagation. They argued that for a relatively small number of collisions, the wave is weakly damped because initially neutrals do not participate in the wave motion and do not share the same momentum. However, for much larger collision frequencies, the plasma species drag the neutrals along, and all three components (electrons, ions, and neutrals) move together. For stronger collisions, Vranjes et al. suggested that the propagations lengths will be large. Our experimental observations suggest that we may have hit this parametric space.

Table 1: Plasma parameters measured

| Pressure (mbar) | $n_n$ ($m^{-3}$) | $\omega/\omega_{pi}$ | $k\lambda_{de}$ | $\nu_{en}$ (Hz) | $\nu_{in}$ (Hz) | $\lambda_{en}$ (cm) | $\lambda_{in}$ (cm) | $\lambda_{wave}$ (cm) | $T_e$ (eV) |
|---|---|---|---|---|---|---|---|---|---|
| $2 \times 10^{-4}$ | $4.8 \times 10^{18}$ | 0.015 | 0.045 | $1.5 \times 10^6$ | $3 \times 10^3$ | 62 | 25.56 | 6.7 | 5.2 |
| $5 \times 10^{-4}$ | $1.2 \times 10^{19}$ | 0.0092 | 0.025 | $3.4 \times 10^6$ | $6.8 \times 10^3$ | 25 | 10.22 | 5.5 | 4.0 |
| $10 \times 10^{-4}$ | $2.4 \times 10^{19}$ | 0.0039 | 0.017 | $5.7 \times 10^6$ | $1.1 \times 10^4$ | 12 | 5.11 | 4.2 | 3.0 |
| $20 \times 10^{-4}$ | $4.8 \times 10^{19}$ | 0.0030 | 0.014 | $9.9 \times 10^6$ | $2 \times 10^4$ | 6 | 2.55 | 3.1 | 2.2 |

The parameters described in table 1 are discussed as follows: electron Debye length, $\lambda_{de}$ is calculated using electron temperature, $T_e$, and the ion plasma frequency, $\omega_{pi}$ is calculated using the plasma density as shown in figure 2, $n_n$ is neutral density, ion temperature $T_i = T_e/10$, ion-neutral collision cross section $\sigma_{in} = 80 \times 10^{-20}\ m^{-2}$, electron-neutral collision cross section $\sigma_{en} = 3.3 \times 10^{-19}\ m^{-2}$ [47], $\lambda_{en} = 1/n_n\sigma_{en}$, $\lambda_{in} = 1/n_n\sigma_{in}$, $\nu_{en} = n_n\sigma_{en}V_{the}$, and $\nu_{in} = n_n\sigma_{in}V_{thi}$ are electron-neutral and ion-neutral collision mean free path and collision frequencies respectively. The experiments are performed for a constant frequency, $f = 100$ kHz, (wave frequency, $\omega = 2\pi f$).

Table 1 indicates that, IA wave frequency is greater that the ion-neutral collision frequencies, for all the values of neutral pressures at which experiments are performed. It shows that for present work the IA wave experiments are performed for $\omega > \nu_{in}$ regime. Therefore, to understand the experimental observation of enhancement in IA wave propagation length, we look for the dynamics related to the electron-neutral, ion-neutral collision mean free path ($\lambda_{en}$ and $\lambda_{in}$ respectively) and wavelength of IA wave ($\lambda_{wave}$). Ions and neutrals are comparable in mass and size and electrons are small comparatively, so probability of momentum transfer from ions to neutrals is more than electrons. As one can observe from the table 1 that for low neutral density, $\lambda_{wave}$ (6.7 cm) is very small in comparison to $\lambda_{in}$ (25 cm), in this condition only few collisions will take place between ions and neutrals before the ions hit to wall of the device. It suggests that the ions interaction with neutrals is negligible, and as a result ions do not impart the IA wave momentum to neutrals. Hence the neutrals do not participate in the wave motion. However, as the neutral pressure increases, $\lambda_{wave}$ (3.1 cm) becomes comparable to $\lambda_{in}$ (2.5 cm). Collisions between ions and neutrals will take place more frequently, it increases the interaction of ions with neutrals; hence the ions transfers and share the IA wave momentum with neutrals. The plasma species drag the neutrals along, and all three components (electrons, ions, and neutrals) move together. The neutrals also start to carry the IA wave momentum as ions and electrons do, and it leads to increase in the wave propagation length as the neutral pressure increases.

The above discussion qualitatively can be understand using the analytical work of Doucet et al. We have extended their solution for IA wave to include collisions between ions and neutrals and the obtained relation for perturbed density ($n_1$) as:

$$n_1 = \lambda \sqrt{n_0} \cos kz$$

where,

$$k = \sqrt{k_0^2 + ik_0 \frac{\nu_{in}}{c_s} - \frac{1}{\sqrt{n_0}} \frac{d^2(n_0^{1/2})}{dx^2}}$$

Here, $n_0$- mean plasma density, $k_0 = \omega/c_s$, $\omega$- IA wave frequency, $c_s = \sqrt{k_B T_e/m_i}$, $k$- wavenumber, $\nu_{in}$- ion-neutral collision frequency, $\lambda$- arbitrary constant. The detail of the calculation can be found in Appendix A.

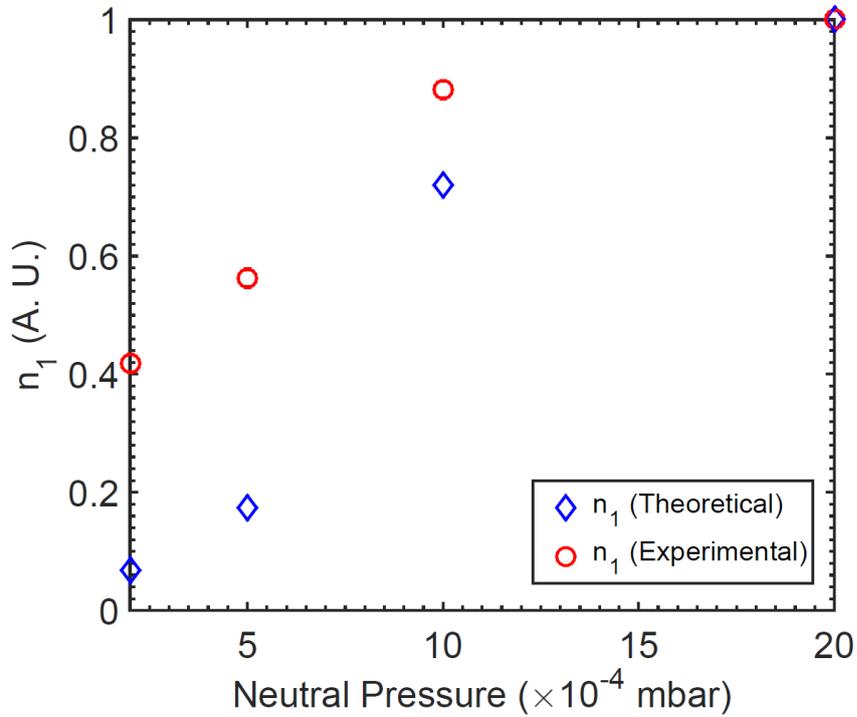

*Figure 10: Variation of perturbed density amplitude ($n_1$) with neutral pressure, experimental values are shown by circle and theoretically calculated values using Eq(1) are shown by diamond marker.*

The values of $k$ are calculated by substituting the experimentally obtained parameters and real part of k is used to calculate the perturbed density. Perturbed density amplitude measured experimentally and calculated theoretically shows qualitative match as shown in Figure 10.

## VI. Summary and Discussion

Ion-acoustic wave has been excited in Argon plasma driven by the localized grid, and the basic features of IA wave in plasma have been studied and discussed. The observation of IA wave excitation has been characterized further and validated. To explore the effect of the neutral collision on IA wave the experiment has been performed for various neutral densities. The neutral density is varied by changing the neutral pressure, which in turn allows the change in ion-neutral, and electron-neutral collision mean free path. It is obvious and conventional that as the neutral collision increases the IA wave get heavily damp in plasma. Our experimental observation shows rather contrary results: it shows that, the wave amplitude ($\delta n/n$) and observable propagation length of IA wave enhances as the neutral density increases. We observe that collisions support the wave to propagate for larger distances as the neutral density increases. This phenomenon is not often observed in IA wave experiments. Increase in neutral density leads to decrease in ion-neutral collision mean path. For the experiments represented here, the ion-neutral mean free path becomes comparable to the wavelength of IA wave and ions impart the momentum of IA wave to neutrals. The collisions of electrons and ions make the neutrals to carry and share the IA wave momentum. It makes the wave to survive and propagate for larger distances in plasma. Experimental findings are shown to be qualitatively consistent with the estimates from a simple fluid model of collisional IA wave.

## Acknowledgments

The authors would like to express the sincere thanks to Minsha Shah and Pankaj Srivastav for their help in the electronic circuit development and Dr. G. Ravi for critical review of the manuscript.

## Appendix A

In the following, we extended the calculations of Doucet et al. to include the effect of neutral collisions. Momentum and continuity equation for plasma species electron and ions (s=e, i) in presence of collisions and absence of magnetic field are written as:

$$n_s m_s \left(\frac{\partial v_s}{\partial t} + v_s \cdot (\nabla v_s)\right) = n_s q_s E - \nabla p_s - n_s m_s \nu_{sn} v_s$$

$$\frac{\partial n_s}{\partial t} + \nabla \cdot (n_s v_s) = 0$$

where, $\nabla p_s = k_B T_s \nabla n_s$,

The following assumptions are made:

1. $T_e > T_i$, $\nabla p_i$ can be neglected comparison to $\nabla p_e$.
2. $\omega \ll \omega_{pe}$, hence the electron inertia and $\nu_{en}$, electron neutral collisions are neglected.
3. $n_e = n_i = n$, as the plasma is quasi-neutral.

The plasma parameters can be written as

$n = n_0 + n_1, E = E_0 + E_1, v_s = v_{s0} + v_{s1}$, where $n_1 \ll n_0, E_0 = 0, v_{i1} = v_{e1}$

The sinusoidal oscillating perturbed quantity can be written as:

$$A_1 = \widehat{A_1} e^{i(kz - \omega t)}$$

Linearizing the equations using the plasma approximations, we obtain the following

The electron momentum equation is,

$$0 = n_0 q \nabla \phi_1 - k_B T_e \nabla n_1$$

The ion momentum equation is,

$$n_0 m_i \left(\frac{\partial v_1}{\partial t}\right) = -n_0 q \nabla \phi_1 - n_0 m_i \nu_{in} v_1$$

and continuity equation is,

$$\frac{\partial n_1}{\partial t} + \nabla \cdot (n_0 v_1) = 0$$

The solution of above three equations gives

$$n_1(z,t) = n_0(z)\frac{e\phi_1(z,t)}{k_B T_e}$$

$$\frac{\partial v_1}{\partial t} = -\frac{e}{m_i}\frac{\partial \phi_1(z,t)}{\partial z} - v_1 \nu_{in}$$

$$i\omega \frac{n_1(z,t)}{n_0(z)} - \frac{\partial v_1}{\partial z} - v_1 \frac{1}{n_0(z)}\frac{\partial n_0(z)}{\partial z} = 0$$

Solving the above three equations we have,

$$\frac{\partial^2 u}{\partial z^2} + h\frac{\partial u}{\partial z} + \frac{\omega^2}{c_s^2}u + i\frac{\omega \nu_{in}}{c_s^2}u = 0$$

$u = n_1(z)/n_0(z)$, $h = (\partial n_0/\partial z)/n_0 = 1/L_n(z)$,

To solve the above second order differential equation we assume,

$u = y(z)\cos kz$,

$u' = y'\cos kz - ky\sin kz$,

$u'' = y''\cos kz - ky'\sin kz - ky'\sin kz + k^2 y\cos kz$,

Substituting back in the equation, we have

$$\left(y'' + hy' - k^2 y + k_0^2 y + ik_0\frac{\nu_{in}}{c_s}y\right)\cos kz - k(2y' + hy)\sin kz = 0$$

Sufficient conditions for above equation to be satisfied for all values of $x$ are that,

$$2y' + hy = 0$$

and

$$y'' + hy' - k^2 y + k_0^2 y + ik_0\frac{\nu_{in}}{c_s}y = 0$$

$$2\frac{1}{y}\frac{dy}{dz} + \frac{1}{n_0}\frac{dn_0}{dz} = 0$$

Where, $h = \frac{1}{n_0}\frac{dn_0}{dz}$, it gives

$$y(z) = \frac{\lambda}{\sqrt{n_0(z)}}$$

$\lambda$ is arbitrary constant.

$$y' = -\frac{1}{2}\frac{\lambda}{n_0^{3/2}} n_0'$$

$$y'' = -\frac{1}{2}\frac{\lambda}{n_0^{3/2}} n_0'' + \frac{3}{4}\frac{\lambda}{n_0^{5/2}} (n_0')^2$$

Substituting these values of $y, y', y''$ into the equation, we have

$$\left(-\frac{1}{2n_0} n_0'' + \frac{3}{4n_0^2} (n_0')^2 - \frac{h}{2n_0} n_0' + k_0^2 + ik_0 \frac{v_{in}}{c_s}\right)\frac{\lambda}{\sqrt{n_0}} = k^2 \frac{\lambda}{\sqrt{n_0}}$$

$$k^2 = k_0^2 + ik_0 \frac{v_{in}}{c_s} - \frac{1}{2n_0} n_0'' + \frac{3}{4n_0^2} (n_0')^2 - \frac{1}{2n_0^2} (n_0')^2$$

$$k^2 = k_0^2 + ik_o \frac{v_{in}}{c_s} - \frac{1}{2n_0} n_0'' + \frac{1}{4n_0^2} (n_0')^2$$

$$k^2 = k_0^2 + ik_0 \frac{v_{in}}{c_s} - \frac{1}{\sqrt{n_0}}\frac{d^2(n_0^{1/2})}{dz^2}$$

As we assumed $u = y(z) \cos kz$, and $u = n_1/n_0$, so $n_1$ becomes

$$n_1 = \lambda\sqrt{n_0} \cos kz$$

where,

$$k = \sqrt{k_0^2 + ik_0 \frac{v_{in}}{c_s} - \frac{1}{\sqrt{n_0}}\frac{d^2(n_0^{1/2})}{dx^2}}$$

where, $k_0 = \omega/c_s$ and $k$ is wave number, $c_s = \sqrt{k_B T_e/m_i}$